\documentclass[a4paper,12pt]{article}
\usepackage{graphicx,epsf,a4wide}
\newcommand{\eq}[1]{Eq.~(\ref{#1})}
\author{G. Bonnet and F. David\\
        \smallskip
        CEA/Saclay, Service de Physique Th\'eorique\\
        F-91191 Gif-sur-Yvette Cedex, France
}
\title{Renormalization Group for Matrix Models with Branching Interactions}
\begin{document}
\begin{titlepage}
\raisebox{-1.2cm}{\includegraphics[width=5.cm]{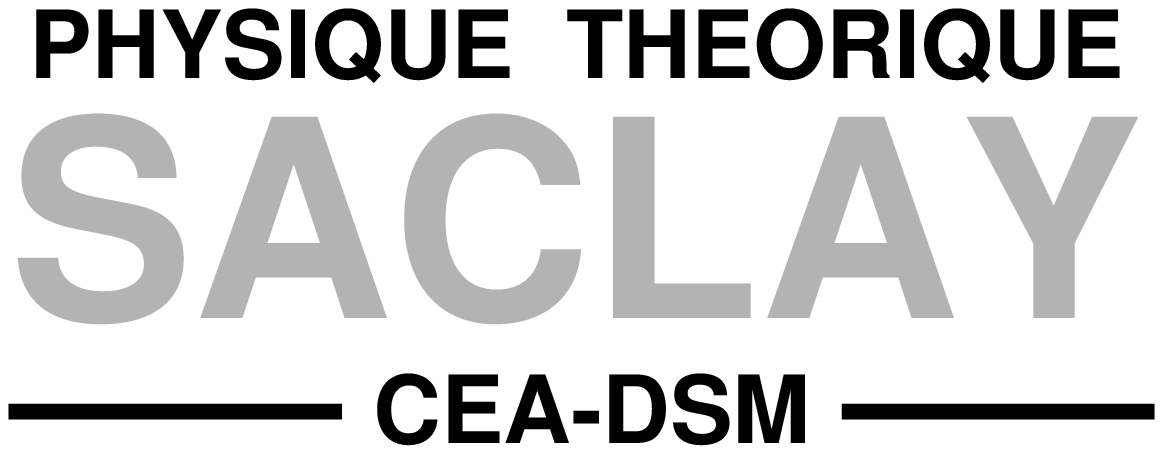}}
\hfill
\begin{minipage}[t]{3.cm}
{\mbox{hep-th/9811216}\\ Saclay T98/123\\ revised version\\}
\end{minipage}
\\[1.cm]
\begin{center}
\textbf{\LARGE Renormalization Group for Matrix Models\\\medskip with Branching Interactions}
\\[4.ex]
{\large Gabrielle \textbf{Bonnet}\,\footnote{AMN}
\footnote{gabonnet@spht.saclay.cea.fr} and 
Fran\c cois \textbf{David}\,\footnote{Physique Th\'eorique CNRS}
\footnote{david@spht.saclay.cea.fr}}
\\
        \bigskip
        CEA/Saclay, Service de Physique Th\'eorique\\
        F-91191 Gif-sur-Yvette Cedex, France
\end{center}
\bigskip

\begin{abstract}
We develop a method to obtain the large $N$ renormalization group flows for
matrix models of 2 dimensional gravity plus branched polymers.
This method gives precise results for the critical points and exponents for one
matrix models.
We show that it can be generalized to two matrices models and we recover the
Ising critical points.
\end{abstract}
\end{titlepage}%
\section{Introduction:}
\label{intro}

Random matrices are useful for a wide range of physical problems.
In particular, by the means of Feynman rules, random matrices can be
interpreted in term of two-dimensional surfaces, which themselves are related
to two-dimensional quantum gravity \cite{refgen}.
2D quantum gravity models can be coupled to matter fields, with a non-zero
central charge $C$.
In term of matrices, this leads to consider 
multi-matrix models, which, studied near their critical points, in the
scaling limit, allow to recover a continuous theory.

Some of the random matrices models have been solved exactly \cite{BIPZ} and
the continuous limit one recovers in the neighborhood of their critical points
has been related to $C\leq 1$ quantum gravity models.
Although the behaviour of $C \leq 1$ models is well understood, exact methods
have not been able to solve $C>1$ models yet.
There is a ``$C=1$ barrier'', which prevents us from using $C>1$ matrix
models as a tool to understand $C>1$ two-dimensional quantum gravity.

Whereas exact technics are unable to deal with $C>1$ models, an
approximate method is likely to succeed better.
In 1992, Brezin and Zinn-Justin \cite{BrezinZinn92} introduced a new method to solve
random matrices models: the large $N$ renormalization group (RG), where
the rescaling parameter $N$ is the size of the matrices.
Integrating over part of the matrices to reduce $N$ one obtains a RG flow in
the space of actions.
Fixed points should correspond to critical points of matrix models in the large
$N$ limit and the scaling dimensions of operators give the corresponding
critical exponents.
For instance, the scaling dimension $\lambda_0$ of the most relevant operator
is related to the ``string exponent'' $\gamma_s$ by the relation
$\gamma_s=2 -{2 \over \lambda_0}$

While the large-$N$ renormalization group method was introduced to study 
directly $C>1$ models, its application to already solved (by exact methods)
$C \leq 1$ models is also useful.
Indeed, in \cite{FD} it was argued by one of the authors that
the understanding of the behaviour of the flows for $C \leq 1$ models
would throw light on what happens at $C>1$:
taking into account ``branching interactions'' in matrix models, it is expected
that in addition to the gravity fixed point there is another fixed point
(corresponding to the ``branching" transition between 2D gravity and
branched polymer behavior).
The value of the critical exponents when $C \rightarrow 1$ (known from
the exact solutions and KPZ scaling) support the conjecture of \cite{FD} that
at $C=1$ these two fixed points would merge,
and that there would be generically no gravity fixed point for $C>1$ models.

In this paper we develop the large $N$ RG method in order to study 
precisely matrix models for 2D gravity + branched polymers.
We shall describe first in Sec.~\ref{RGMM} the renormalization group method as 
it was introduced by Brezin and Zinn-Justin, and the improvements made by
Higuchi et al. \cite{Higuchi et al}, using the equation of motions.
Then we explain why these methods are not sufficient to study our class 
of models.
We show in Sec.~\ref{OM} that the linear approximation of \cite{Higuchi et al}
is not sufficient and we propose a method to go further, and we apply it to the 
one matrix model.
The method still requires numerical analysis, which is performed in Sec.~\ref{NA},
where the resulting RG flow equations are analysed and the results of 
the method are compared to the previous results of \cite{Higuchi et al} 
and to the exact results.
Finally in Sec.~\ref{IsingSection} we shall consider the generalizations of our
method, in particular to the Ising plus branched polymer two-matrix model. 
 
\section{The Renormalization Group for Matrix Model}
\label{RGMM}
\subsection{General Idea}
We shall consider a matrix model for gravity plus branched polymers.
The partition function is
\begin{equation}
Z\ =\ \int d\Phi \ e^{-N^2 V(\Phi)} \hspace{20pt} \Phi 
\makebox{ hermitean matrix }
N \times N
\label{defZ}
\end{equation}
where 
\begin{equation}
V(\Phi)\ =\ T_2 + g T_4 + {h \over 2} {T_2}^2\qquad\makebox{and}\qquad 
T_n\ =\ {1\over N}\ \makebox{tr} \left[ {\Phi^n \over n} \right]
\label{VandTn}
\end{equation}
In \eq{VandTn} $g$ is the gravity coupling constant and $h$ is 
the branched polymer coupling constant.
This model can be solved exactly, thus it should be a
good test of renormalization group methods.
When $h$ is equal to zero, it is the pure
gravity model that was considered by Brezin and Zinn-Justin in 
\cite{BrezinZinn92}
when they first tested their renormalization group (RG) method.
The general idea of RG is to start from the potential $V(\Phi')$ for a
$N'\times N'$ matrix $\Phi'$ with $N'=N+1$,
and to integrate over the last row and the last column of
the matrix, leading to an effective action $V'$ for the $N\times N$
remaining matrix $\Phi$.
Performing a linear rescaling of $\Phi$
(wave-function renormalization) to keep the $\mathrm{tr}(\Phi^2)$ term
unchanged one obtains the renormalized action 
$V(\Phi)+N^{-1}\Delta V(\Phi)$ and the RG flow equation in the space of
action ${\scriptstyle N}{\partial\over\partial N}V=\Delta V$.

In \cite{BrezinZinn92} it was shown that at order
$g^3$ the action (\ref{VandTn}) (with $h=0$) stayed closed (up to a additive shift coresponding to a $\mathrm{tr}(\mathbf{1})$ operator), i.e.
no new operators were generated by the RG.
The corresponding beta-function for $g$ was found to have a zero for
$g<0$, corresponding to the critical coupling where pure gravity scaling
is recovered.
However at that order the first numerical result of \cite{BrezinZinn92}
were only very qualitative, since
the error on the critical coupling itself is of order 100\%.
This method was then developped further by several authors \cite{various93}.

Shortly after, Higuchi, Itoi, Nishigaki and Sakai \cite{Higuchi et al}
improved the RG method by using the equations of motion of the matrix model
(the so-called loop equations) to eliminate some of the new operators
which appear in the RG transformation at higher orders in $g$.
They first considered the pure gravity model with cubic action
\begin{equation}
V(\Phi)=T_2+g T_3
\label{IguchiV}
\end{equation}
and showed that in the planar limit, the RG equation for the
free energy $F(g)$ can be written as 
\begin{equation}
\label{linearapprox}
N {\partial F \over \partial N}+
2 F =G(g,{\partial F \over \partial g})
\end{equation}
where $G$ is a non-linear function of $g$, the gravity coupling constant, and
of ${\partial F \over \partial g}=\langle T_3\rangle$.
This is a non linear differential equation, at variance with the standard RG
equations which should be linear and of the generic form given in \cite{BrezinZinn92}
\begin{equation}
N {\partial F \over \partial N} -\beta(g) {\partial F \over \partial g}
+ \gamma(g) F= r(g)
\end{equation}
(where $\gamma(g)$ and $r(g)$ are regular functions in $g$).
In order to obtain a standard renormalization group equation the authors of \cite{Higuchi et al} truncated $G(g,{\partial F \over \partial g})$ to
the first order in $\partial F \over \partial g$.
For this problem this approximation is in fact quite good. 
The results they obtained for the cubic action (\ref{IguchiV})
are : a $0.2 \%$ error for the value of the critical
coupling and a $2.5\%$ error for the eigenvalue corresponding to the string
critical exponent $\gamma$.
Moreover, they have  generalized this method to a one-matrix model with
two couplings: $g_3 T_3 +g_4 T_4 +T_2 $ and also to the Ising two-matrix model,
with in both cases results quite close to the exact results.
In particular for the quartic action ($g_3=0$) the relative errors on the
coupling and the eigenvalue are $1\%$ and $5\%$.

Nevertheless, the linear approximation method has only been
applied up to now to gravity models without branching interactions.
The case of gravity plus branched polymers, which are the models one has to 
consider if one wants to verify the scenario of \cite{FD}, would be more 
difficult to treat.
One would have to introduce ${\partial F \over \partial h}$, partial
derivative of $F$ with respect to $h$, and linearly develop the
equations in ${\partial F \over \partial h}$ and ${\partial F \over
\partial g}$.
Moreover, the success of the linear approximation method of
 \cite{Higuchi et al} may
be attributed to the fact that the value of the critical coupling for
pure gravity, $g_c\simeq -0.22$, is in fact quite small, since
the corrections to the linear approximation turn out to be of order $g^3$,
and are therefore small.
For branching interactions, we expect that this will not be the case.
Already for the pure branched polymer model ($g=0$, $x<0$) the critical point is $h_c=-0.5$, thus we can guess 
that the estimates for the general critical points would be far less
precise than those for the pure gravity critical point.

\subsection{The Equations of Motion :}
Since in this paper we shall also use the equations of motion to write
RG equations for matrix models with branching interactions, let us 
briefly recall the general idea.
If one starts from a simple action $V$ such as (\ref{VandTn})
(which depends only on the two operators $T_2$ and $T_4$)
one obtains a variation $\Delta V$ which is a function of all the $T_{2 n}$ 
for any $n$.
To take all these terms into
account, one can try to write the action of the renormalization group
on the most general partition function.
This leads to expressions 
with an infinite number of terms which must be truncated in some way.
However we know no general and natural truncation scheme, and the simplest
truncations lead to a very slow 
convergence for the value of critical points, while the estimates for the
critical exponents do not converge at all!
This can be shown explicitely on the simple example of the pure branched
polymer model with action $ V(\Phi)=T_2+{h \over 2} {T_2}^2$ (which is in fact
a vector model) as discussed in Appendix A.

Nevertheless one must realize that all the operators are not independent.
One is free (in perturbation theory) to include in the RG transformation
a non-linear reparametrization of the field variable (i.e. the matrix $\Phi$) 
in the partition function, of the form
\begin{equation}
\Phi\ \to\ \Phi + \epsilon(\Phi)\qquad,\qquad\epsilon(\Phi)\ =\ N^{-1}\,\sum_k A_k[\{T_j\}]\Phi^k
\label{}
\end{equation}
This induces an additional variation in the effective action of the form
\begin{equation}
\delta V\ =\ \epsilon(\Phi)\cdot{\partial V\over\partial\Phi}-
{\partial \epsilon(\Phi)\over\partial\Phi}
\end{equation}
This variation does not change the physical content of the RG equations, since
$\langle\delta V\rangle=0$ 
where $\langle\cdots\rangle$ denotes the ``vacuum expectation value'' 
(v.e.v.)  
$Z^{-1} \int d \Phi (\cdots) e^{-N^2 V(\Phi)}$;
these are the quantum equations of motion of the model, which express the
relations between the v.e.v. of the traces.
This (unphysical) arbitrariness in the RG equations is known as the problem
of the ``redundant operators''\cite{Wegner76}.
The idea is that one should use the equation of motion to simplify the
expression of the RG equations and to reduce the number of operators and
of coupling constants involved in the RG flows.
This is what has been done in \cite{Higuchi et al} 
to reduce the RG equation to the form
of \eq{linearapprox}.
In Appendix \ref{VectorAppendix} we show how the method applies to the simple branched polymer
model.

We are now going to introduce our method, which aims at
calculating the renormalization flows of gravity plus branched polymer
models. 

\section{Our Method :}
\label{OM} 

\subsection{The Action of the Renormalization Group :}

We are now going to explain our method on the example of the one matrix model :
$V(\Phi)=g T_4+ T_2+ {h \over 2} T_2^2 $ (More general cases as the Ising plus
branched polymer model will be considered in section \ref{IsingSection} and we shall
discuss then the generalization of our method).
As the use of the equations of motion does not allow us to put the 
renormalized action exactly in the same form as the action we would like
to study, we are going to consider a slightly more general model :
\begin{equation}
V(\Phi)\ =\ \psi(T_2)\ +\ g\, T_4
\label{Vt4}
\end{equation}
where $\psi$ is of the form :
\begin{equation}
\psi\ =\ \sum_{j=1}^{\infty} 2^{-(j-1)} h_j T_2^j
\label{psii}
\end{equation}
with $h_1=1$ and $h_2=h$.
We are going to show that when the renormalization group acts on this
model, the renormalized action can be put in the same form, with a 
renormalized $\psi$ and a renormalized $g$, and $h_1$ always equal to one.

In the following we denote $U$ the derivative of $\psi$
\begin{equation}
{\partial \psi \over \partial T_2}=U(T_2)
\label{Udef}
\end{equation} 
We start from $\Phi'= \left( \begin{array}{cc}
\Phi & v\\
v^{\ast} & \alpha\\
\end{array} \right)$
a $(N+1) \times (N+1)$ hermitian matrix, $\Phi$ is a $N \times N$ hermitian 
matrix, $v$ is a vertical vector, and $\alpha$ a real number.
\begin{equation}
Z_{N+1}\ =\ 
 \int d \Phi' \ \  \exp\left[-(N\ +\ 1)^2 \left(g {\makebox{tr} 
\Phi'^4 \over 4 
(N\ +\ 1)}
\ +\  \psi\left({\makebox{tr} {\Phi'}^2 \over 2 (N\ +\ 1)}\right)\right)\right]
\label{znplus1}
\end{equation}
can be rewritten, to the first order in ${1 \over N}$, separating the variables $\Phi$, $v$, $v^*$ and $\alpha$,
 as :
\begin{eqnarray}
\int \  d\Phi\  dv\   dv^{\ast}\  d\alpha\  
\exp\left[-N^2\ \left(V(\Phi)\ +\ {1 \over N}
\Delta V(\Phi)\right)\right] \qquad \makebox{with }\qquad \nonumber\\
\Delta V= g T_4 + 2 \psi  - T_2 U  + g {\alpha^4 \over 4} + U 
{\alpha^2 \over 2}  + 
v^{\ast}(U + g \alpha^2  + g \alpha  \Phi  + g \Phi^2) v  + g
 {(v^{\ast}  v)^2 \over 2}  
\end{eqnarray}
Let us introduce first the auxiliary field $\sigma$ and rewrite the part of 
the integral containing the variables $v$ and $v^{\ast}$ as :
\begin{equation}
I=\int dv dv^{\ast}  \exp\left[-N (v^{\ast} (U+g \alpha^2+g \alpha  \Phi+g \Phi^2+g \sigma)v
-g {\sigma^2 \over 2})\right]
\label{i}
\end{equation}

By integrating over $v$ and $v^*$, we obtain,

\begin{equation}
I=\exp[-N \left(-g {\sigma^2 \over 2}+
{1 \over N} \makebox{tr} \left[\ln(U+g \alpha^2+g \alpha \Phi+g \Phi^2+g
\sigma)\right] \right)]
\label{ibis}
\end{equation}
We then can use a saddle point method and minimize this expression with
 respect to $\sigma$. We find an implicit
equation for the value $\sigma_s$ of $\sigma$ :
\begin{equation}
g \sigma_s \ =\ \left({1 \over N}\right)
\makebox{tr}\left[{g \over U\ +\ g \sigma_s \ +\ g \Phi^2\ +\ g \alpha \Phi\ +\ g 
\alpha^2}\right]
\label{sigma}
\end{equation}
We now have to
integrate over $\alpha$.
This can also be done by  using the saddle point method.
The ${d \sigma_s \over d \alpha_s}$ terms disappear thanks to \eq{sigma}, and
the saddle point $\alpha_s$ verifies :
\begin{equation}
g \alpha_s^3 \ +\ U \alpha_s \ +\ {1 \over N} \makebox{tr}\left[ {2 g \alpha_s
\ +\  g \Phi
\over U\ +\ g \alpha_s^2 \ +\ g \alpha_s \Phi\ +\ g \sigma_s}\right] \ = \  0  
\label{alpha}
\end{equation}
Then we can rewrite, where $\langle \cdots \rangle$ means that the average is done over $\Phi$,
\begin{equation}
 {Z_{N+1} \over Z_N}\ =
 \langle \exp\left[-N\left( g T_4 + 2 \psi  - T_2 U 
 + g { \alpha_s^4 \over 4} + g
 { \alpha_s^2 \over 2} + A - g { \sigma_s^2 \over 2}\right)\right]\rangle 
\label{moy}
\end{equation}
with
\begin{equation}
 \ A\ =\ \left({1 \over N}\right) \makebox{tr} \left[ \ln(U + g \alpha_s^2
 + g \alpha_s 
\Phi +g \Phi^2 +g \sigma_s)\right] 
\end{equation}
and use the factorization property which is valid in the large $N$ limit
$\langle A B \rangle=\langle A\rangle \langle B\rangle +O(N^{-2})$
 to express the variation of $V$:
\begin{eqnarray}
\langle N {\partial V \over \partial N}\rangle \  =\ \langle
 g T_4\ +\ 2 \psi\ -\ T_2 U\ +\ g {{\langle\alpha_s\rangle}^4 \over 4}
\ +\ g {{\langle\alpha_s\rangle}^2 \over 2}\ +\nonumber\\
+\ {1 \over N} \makebox{tr}\left[ \ln(U\ +\ g {\langle\alpha_s\rangle}^2\ +\ 
g \langle\alpha_s\rangle \Phi \ +\ g \Phi^2 \ +\ g \langle\sigma_s\rangle)\right]
\ -\ g {{\langle\sigma_s\rangle}^2 \over 2}\rangle
\label{reneq}
\end{eqnarray}
From the saddle point equation on $\alpha_s$, and the parity of the action 
which leads to $\langle\makebox{tr}\Phi\rangle\ =\ 0$ we immediately see
that $\langle\alpha_s\rangle\ =\ 0$ is
solution of the averaged saddle point equation.

Thus :
\begin{equation}
 \langle -N {\partial V \over \partial N}\  =\ 
g T_4 \ +\ 2 \psi \ -\ T_2 U \ +\ {1 \over N}
\makebox{tr}\left[ \ln (U\ +\ g \sigma_s \ +\ g \Phi^2)\right]
\ -\ g {\sigma_s^2 \over 2}\rangle 
\label{reneqbis}
\end{equation}
\begin{equation}
\makebox{with }\qquad  \langle\sigma_s\rangle\ =\ \langle{1 \over N} \makebox{tr}
 (U\ +\ g \sigma_s\ +\ g \Phi^2)^{-1}\rangle
\label{sigbis}
\end{equation}

\subsection{The Equations of Motion :}

As discussed in Sec. 2.2, one should use the equation of motion and the freedom
to add redundant operators to the RG transformation to simplify the flows.
We now express the equations of motion.
The change in variables :
$\Phi \rightarrow \Phi+\epsilon \Phi^{2 n +1}$, $n \geq 0$
leads to the equation of motion :
\begin{equation}
\langle((n\ +\ 2)\, g\, T_{2 (n+2)}\ +\ (n\ +\ 1) \, U \, T_{2 (n+1)} 
\ -\ 2 \sum_{i=0}^{n}
i \, (n\ -\ i)\, T_{2 i}\, T_{2 (n-i)}\rangle\ =\ 0
\label{eqmvt1}
\end{equation}
Where the $T_n$ are defined as in \eq{VandTn}
\begin{equation}
{\displaystyle n T_n ={1 \over N}\makebox{tr} \left(\Phi^n \right)}
\label{Tnagain}
\end{equation}
We introduce the function $f(z)$ which is the generating function of the
$\langle T_{2n} \rangle$, $f(z)={1 \over N}\langle\makebox{tr} {1 \over 1-z \Phi^2 }\rangle$
(this function $f$ is almost the
resolvent of the model). Using \eq{eqmvt1} $f(z)$ can be rewritten :
\begin{equation}
f(z)\ =\ \langle {(g\ +\ U z)\ +\ \sqrt{(g\ +\ U z)^2 \ -\ 4 z^2 (g\ +\ U z\ +\ 
2 g T_2 z)} \over 2 z^2} \rangle
\label{f}
\end{equation}

Noting that 
\begin{equation}
\langle\sigma_s\rangle=
\langle{1 \over N} \makebox{tr}\left[(U+g \sigma_s +g \Phi^2)^{-1}\right]\rangle
=
{1 \over (U+g \sigma_s) } f\left({-g \over U+g \sigma_s}\right)
\label{vevsigma}
\end{equation}
we immediately find that the solution of the above equation is
\begin{equation}
\langle \sigma_s \rangle = 2  \langle T_2 \rangle
\label{solsig}
\end{equation}
Integrating $f$, we also have
\begin{equation}
\langle{1 \over N}\,\makebox{tr}\left[\ln (U + g \sigma_s + g \Phi^2)\right]
\rangle
\ =\ \langle\ln(U + g \sigma_s)\ +\ 
\int_0^{-{g \over U + g \sigma_s}} {f(z) - 1 \over z}\ dz\rangle
\label{tralint}
\end{equation}
Finally, denoting
\begin{equation}
\rho =-{g \over U^2} \qquad,\qquad x=2 T_2 U
\label{rhoandx}
\end{equation}
we obtain
\begin{eqnarray}
\langle -N {\partial V \over \partial N}\ =\ g T_4\ +\ 2 \psi \ +\ \ln U \ -\ 
{x \over 2}\ +\ 
\ln(1\ -\ \rho x)\ +\ \rho {x^2 \over 2} \ + \nonumber\\
+\int_0^{\rho \over 1 - \rho x} dz\ {z - \rho - 2 z^2 + 
\sqrt{(z - \rho)^2 - 4 z^2 (z
-\rho  - \rho z x)} \over -2 z^3 } \rangle
\label{equafin}
\end{eqnarray}
This expression, if expanded in powers of $T_2$, contains a linear term
in $T_2$ : ${\scriptstyle N} {\partial h_1 \over \partial N}$ (we recall that $\psi(T_2)=
\sum_{j=1}^{\infty} 2^{1-j} h_j {T_2}^j $). We would like to keep $h_1$
equal to one. This is possible if one substracts the coefficient of $T_2$,
multiplicated by the expression appearing in the first equation of motion :
$4 g T_4+2 T_2 U -1$.
Then we finally obtain a renormalized action exactly in the same form as the
original action.

Beyond this point, we shall suppress, for simplicity reasons,
 the $\langle \cdots \rangle$ in all our expressions.
Indeed, \eq{equafin} means that, if we replace $N {\partial V \over \partial N}$
in $Z_{N+1}$ by the right-hand part of our equation, the result will be the same.

\section{Numerical analysis :}
\label{NA}
\subsection{Expansion of the integral :}

We now have obtained a renormalization group equation containing
only linear terms in $T_4$ and powers of $T_2$.
So, starting from the action $g T_4 + \psi(T_2)$
(with $\psi(T_2)=T_2+{h \over 2} T_2^2+\ldots$), we can write 
 (at least in principle) the equations for ${\scriptstyle N} {\partial g \over 
\partial N}$ and ${\scriptstyle N} {\partial \psi \over \partial N}$ :
we just have to expand the above expression in powers of $T_2$.
This method, however, leads to very complicated expressions with many
non-trivial integrals depending on the parameter $g$.
To be able to treat our expression easily, it is better to expand it first
in powers of $\rho$ ($\rho$ is of order $g$, as $U=1+h T_2+\ldots$ is of
order one).
The last integral in \eq{equafin}
is then also expanded in powers of $x$, and it remains to
expand $\rho=-{g \over U^2}$ in powers of $T_2$, which
is easy.

One should notice that the expansion in $\rho$ is \textit{not} trivial :
one cannot
simply expand the integrand and the bounds of integration, as this would
lead to divergent expressions.
In fact, the integral
can be expressed as an expansion in $\rho^n$ and $\rho^n \ln \rho$.
What one has to do to expand the last integral properly is to treat separately
the integral between $0$ and $x \rho^{3 \over 2} {(1-\rho x)}^{-1}$, 
and between 
$x \rho^{3 \over 2} {(1-\rho x)}^{-1}$ and $\rho (1-\rho x)^{-1}$.
Let us briefly describe these operations : the last integral can be 
expressed as a sum of two integrals :
\begin{equation}
I\ =\ 
\int_0^{\rho \over 1 - \rho x} \ {z - \rho - 2 z^2  + 
\sqrt{(z - \rho)^2 - 4 z^2 (z
-\rho  - \rho z x)} \over  -2 z^3 }\ dz \ =\ I_1\ +\ I_2
\label{inte12}
\end{equation}
The first one given by 
\begin{equation}
I_1 \ =\ \int_0^{1 \over \sqrt{\rho}}\  x\  B_1 \ du  
\qquad\mathrm{with}\qquad
{\displaystyle z= {\rho^2 x \over 1\, -\, \rho x} u }
\qquad\mathrm{and}
\label{i1}
\end{equation}
% B_1\  =\ \\
\begin{equation}
B_1 = \rho x\,
{\displaystyle {
%(1-\rho x)\rho x(1-u) 
- 2 (1-\rho x u)^2 
+ x (1-\rho x)
\left((1-u)+\sqrt{ (1 - u)^2 + 4 u (1 - \rho x)^2/x}
\right) 
\over - 2 (1 - \rho x u )^3}  }
\nonumber
\end{equation}
and the second one given by
\begin{equation}
I_2=\int_{x \sqrt{\rho}}^1 B_2\  da 
\qquad\mathrm{with}\qquad
{\displaystyle z={\rho \over 1\, -\, \rho x} a }
\label{i2}
\end{equation}
\begin{equation}
B_2\ =\ 
{\displaystyle {
- 2 \rho (1 - a)^2
+ (1 - \rho x) 
\left((
\rho x - a) +
\sqrt{(\rho x - a)^2 + 4 \rho a (1 - a)^2}
\right)
\over
- 2 (1 - a)^3 \rho}} 
\nonumber
\end{equation}
Before expanding the integrand in powers of $\rho$, up to the
order ${\rho}^{2 n}$, we have to notice that, when $\rho$ tends to zero,
the bounds of integration in $I_1$ and $I_2$ tend respectively to $\infty$
and $0$.
Let us take $I_2$ as an example. For $u \sim {1 \over \sqrt{\rho}}$ large,
the terms in the expansion of the integrand are of the form 
$ \rho^{2 n + 1} u^{2 n}$.
This implies that the integration leads
to a term of order $\rho^n$. Thus, the terms in the expression are really
of increasing powers of $\rho$, which justifies our expansion,
keeping in mind that
expanding the integrand up to order $\rho^{2 n}$ amounts to
expand the integral up to order $\rho^n$.

Moreover, when expanding the integrand $B_1$, we obtain expressions of the
form : $P(u) ((1-u)^2+4 u/x)^{-{m \over 2}}$ where $P(u)$ is a polynomial.
The primitives of such terms lead to logarithmic expressions, and
we finally obtain a $\rho^n$ and $\rho^n \ln \rho$ expansion, that is to 
say, in term of the coupling $g$, we have an expansion in $g^n$ \textit{and}
$g^n \ln(g)$.
We would not have realized the existence of these logarithmic
terms if we had used a finite number of equations of motion;
they appear here because, by
the use of the $f(z)$ function, we have used an infinite number
of equations of motion, and expanded \textit{properly} only after.
This phenomenon should not come as a surprise, it was already observed in
\cite{hikami}.

Once the expansion in $g$ has been done, we expand $U$ in powers of $T_2$ and, 
finally, we obtain a quite simple expression with two orders of
expansion : $X$ in $g$ and $Y$ in $T_2$.

After development in $\rho$ at order $X=4$ in $\rho$, for example,
the expression of the integral $I=I_1+I_2$ is :
\begin{equation}
\begin{array}{l}
I\ =\ 
%I_1\ +\ I_2\ =\\
{\displaystyle \rho \left(-{3 \over 2}+ x - \ln(\rho) \right)  +  
 \rho^2 \big( -12 + 8 x -  6 \ln(\rho) + 4 x \ln(\rho)\Big) +}\\ 
{\displaystyle \rho^3 \left(-{665 \over 6}+ 84 x - {19 \over 2 } x^2 +
  {1 \over 6} x^3 -50 \ln(\rho)  + 36 x \ln(\rho)-3 x^2 \ln(\rho)\right) }\\ 
{\displaystyle + \rho^4 \left(- {3437 \over 3} +{2960 \over 3} x- 174 x^2+{4 \over 3} x^3 +
{1 \over 6} x^4 - 490 \ln(\rho)\right. }\\
{\displaystyle + 400 x \ln(\rho) -60 x^2 \ln(\rho)\Big)
+ o(\rho^4)}\\
\end{array}
\label{trall}
\end{equation}

Inserting the expansion for $I$ in \eq{equafin} we obtain the RG flow equations
for the couplings $g$ and $h_j$ ($j\le Y$).
This flow equations can easily be integrated numerically.

%+ \rho^5 \left(- {63924 \over 5} + {37310 \over 3} x - {5955 \over 2} x^2
% + {239 \over 2} x^3 + {5 \over 12} x^4 + {3 \over 20} x^5 + 
% +  {11 \over 4} x^{12}m  -5292 \ln(\rho) \right. \\
%\left  +4900 x \ln(\rho) - 1050 x^2 \ln(\rho) + 30 x^3 \ln(\rho) 
% \right)+ o(\rho^5)\\

\subsection{The shape of the RG flows :}

\begin{figure}[htb]
\begin{center}
\mbox{\epsfxsize=10cm \epsfbox{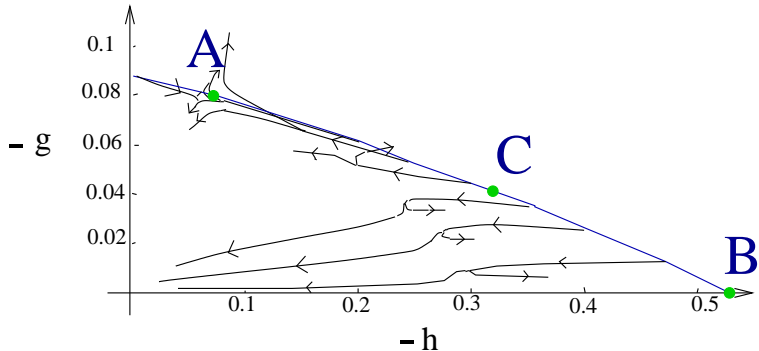}}
\caption{Renormalization flows, where $-h$ is in abcissa and $-g$ in ordinate.
Trajectories may seem to bifurcate : in fact, these are two distinct
 trajectories originating from two points near each other but separated
by the critical line. }
\label{flot1mfig}
\end{center}
\end{figure}
In figure~\ref{flot1mfig} we show  the results obtained at orders
$X=5$ and $Y=6$.
The axes represented correspond to the two couplings that interest us :
the gravity coupling $g$ and the branched polymer coupling $h=h_2$ . 
Of course, there are $Y-2(=4)$ other couplings which are not represented
there.
The RG flows represented here correspond to 
initial condition $h_j=0$, for all the $j>2$, i.e. we study the evolution of
the model with inital action$V=g T_4+T_2+{h \over 2} T_2^2$ as in \eq{VandTn}.
It is easy to find the critical line in the ($g,h$) plane: it separates
the domain where one flows towards the Gaussian fixed point ($g\to 0,h\to 0$)
from the domain where $g$ and $h$ diverge. 
We have chosen here to show only the flows for $(g,h)$ near the
critical line, on both sides of it.
Since under the RG flows the $h_j$'s becomes non-zero, and since we project
the flows on the $(g,h)$ plane the flows may seem to cross, of course this is 
unphysical and a projection of what happens in higher dimension.

First one recovers the correct phase diagram and critical line
(with an average $5\%$ relative error at this order).
The critical line is separated into two parts by a multicritical point
$\mathbf{C}$.
Starting from the rightmost part of the critical line the flows are driven towards a fixed point {\bf B'} which lies on the $g=0$ plane.
Starting from the leftmost part of the critical line the flows are driven
towards a fixed point {\bf A'} (with $g$, $h$, $h_3$, $\ldots$ $\ne 0$).
We have represented on Fig~\ref{flot1mfig} the ``pull-back" {\bf A} of {\bf A'},
i.e. the point on the critical line which flows in the fastest way towards
{\bf A'} (this is equivalent to say that the leading subdominant corrections
to scaling vanish at {\bf A}).
Both fixed points {\bf B'} and {\bf A'} have one unstable direction.
Thus {\bf B'} corresponds to branched polymers and {\bf A'} is the fixed
point corresponding to pure gravity.
Starting from the point {\bf C} one flows towards a fixed point {\bf C'} 
(also with with $g$, $h$, $h_3$, $\ldots$ $\ne 0$) with two unstable directions,
which should therefore correspond to the ``branching transition" between
pure gravity and branched polymer scaling behaviour.
These features of the RG flows are in full agreement with the picture which
was proposed in \cite{FD}.
This agreement is confirmed by studying more precisely the numerical results.

\iffalse
We note immediately that we recover the good critical line (with 
$\simeq 5\%$ error at this order), the bicritical point C (which is
a totally repulsive fixed point of the renormalization group).
We also note the existence of the fixed point A of reference
\cite{} : for this particular model, the shape of the flows fits the
conjecture. For $(g,x)\leq (g_c,x_c)$, the flows are attracted by the 
gaussian fixed point $(0,0)$, and beyond the critical line, they diverge.
%
All these qualitative facts show that, for this particular model,
we have proven 
reference \cite{FD}'s conjecture is right. Of course, to verify the
conjecture further, we have to study the evolution of the two fixed points
(A and C) when the central charge tends to $1$.
\fi

\subsection{Quantitative results :}

At this order of approximation ($X=5$, $Y=5$), we obtain for the critical 
point $g_c$ located on the axis $h=0$ a precision which is smaller than the one obtained by Higuchi et al..
Our computations, however, can be performed for several values of $X$ and 
$Y$, and the results can be extrapolated to obtain a very good precision. 
For example, for $g_c$, critical value of $g$ for $h=0$, we have :\\

\noindent
\begin{tabular}{|c|c|c|c|c|c|}
\hline
$X$ & 2 & 3 & 4 & 5 & $\infty$\\
\hline
$g_c$ & -.095415(5) & -.090745(5) & -.088725(5) & -.087585(5) &-.083325(5)\\
\hline
\end{tabular}

\begin{figure}[htb]
\begin{center}
\mbox{\epsfxsize=10cm \epsfbox{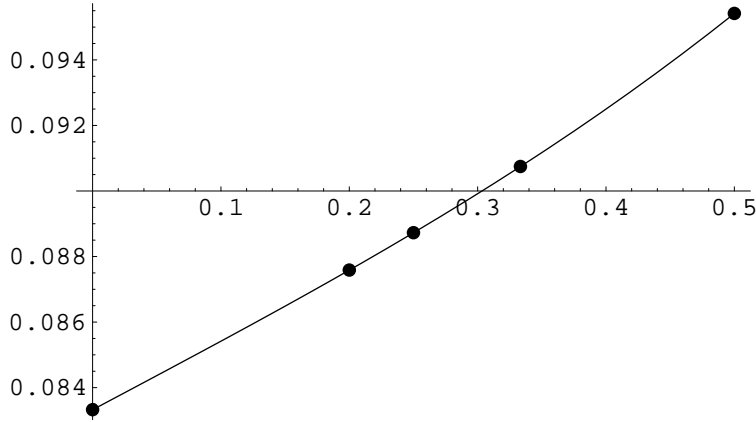}}
\caption{Extrapolating the $h=0$ critical point $-g_c$ as a function of $1/X$}
\end{center}
\label{extrapolgc}
\end{figure}

\bigskip\noindent
We then extrapolate these results by a polynom of degree three in $X^{-1}$.
Fig~\ref{extrapolgc} shows that this
extrapolation behaves almost as a straight line :
the value of $g_c$ converges as $1 \over X$.
The extrapolated value for $X=\infty$ is $-0.083325(5)$, while the exact value
of $g_c$ is $- {1 \over 12} \simeq -0.083333(5)$.
We have thus obtained a $0.016\%$ relative error, to be compared with the 
$0.9\%$ relative error of \cite{Higuchi et al}.\\\

By the same method,
we can obtain the whole critical line in the plane $(g,h)$, which was
not obtained in ref.\cite{Higuchi et al}.
On the pure branched polymer line ($g=0$), we obtain for the branched polymer
critical point
$h_c\simeq -0.50037$ instead of the exact value $h_c=-0.5$ i.e. a $0.07\%$ 
relative error.
For the position of the multicritical point {\bf C} we obtain respectively
for $h_c$ and $g_c$ a $0.02\%$ and $0.5\%$ relative errors.

Linearizing the RG flow equations in the vicinity of the fixed points, we
obtain also good results for the critical exponents.
For instance, studying the multicritical fixed point {\bf C'}, we obtain
for the largest eigenvalue $\lambda_0-1.185$ , i.e. a $1.25\%$ relative error
when compared to the exact result $6/5=1.2$ !

%we have $1.25\%$ error on the eigenvalue $1.2$ corresponding to the critical
%exponent of the bicritical point (this eigenvalue $\lambda$ is related to
%the critical exponent $\gamma$ by the relation : $\gamma=2-{2 \over \lambda}$).
It is not surprising that the critical
exponents converge less rapidly than the positions of the critical points. 
Better but more tedious calculations could be done, for we can go in principle
to any orders $X$ and $Y$, and the limit $X\to\infty$ and $Y\to\infty$ 
should give the exact result, but this is not our purpose here.

\section{Multimatrix Models}\label{IsingSection}

Our method can be generalized to a two-matrix model which describes the
Ising plus branched polymer model : 
\begin{equation}
{\displaystyle V_0=g {\makebox{tr} (A^3 +B^3) \over 3 N}+{\makebox{tr} (A^2 +B^2) \over 2 N}
+{h \over 2} \left( {\makebox{tr} (A^2 +B^2) \over 2 N}\right) ^2 +
 c_0 {\makebox{tr} A B \over N}}
\label{VIsing}
\end{equation}
 $A$ and $B$ are two $N \times N$ hermitian matrices, $g$ is the gravity
coupling constant, $h$ the branched polymer coupling constant and $c$
corresponds to the temperature of the Ising model.
For simplicity reasons, we have not introduced a external field,
which would have broken the symmetry in $A$ and $B$.

As previously, we are led to study a more general model :
if we note $\Phi=\left(
\begin{array}{cc}
A & 0 \\
0 & B \\
\end{array}
\right),  \delta_1=\left(
\begin{array}{cc}
0 & 1 \\
1 & 0 \\
\end{array}
\right)$, and 
$\delta_0=\mathrm{Id}=
\left(
\begin{array}{cc}
1 & 0 \\
0 & 1 \\
\end{array}
\right)$,
the model we are going to study can be expressed as :
\begin{equation}
{\displaystyle V=g {\makebox{tr}\left(\Phi^3\right) \over 3 N}+\psi\left({\makebox{tr} \left(\Phi^2\right)\over 2 N},
{\makebox{tr} \left(\Phi \delta_1 \Phi \delta_1\right) \over 2 N }\right)}
\label{vison}
\end{equation}

Where $\psi$ is, as in the case of one matrix models, a function of the
quadratic terms in $\Phi$.
If 
$T_n={\makebox{tr} \left(\Phi^n\right) \over n N}=
{\makebox{tr}\left(A^n+B^n\right)\over nN}$
and
$T_{AB}={\makebox{tr} \Phi\left(\delta_1 \Phi \delta_1\right) \over 2 N }=
{\makebox{tr}\left(AB\right)\over N}$, we note ${\partial \psi \over 
\partial T_2}
=U$ and ${\partial \psi \over \partial T_{AB}}=c$.
Here $c$ is a series in $T_{AB}$ and $T_2$, with constant coefficient $c_0$.
The renormalization group equation reads :
\begin{equation}
-N {\partial V \over \partial N}=g T_3 +2 \psi-T_2 U -c T_{AB}+g {\alpha^3
\over 3}+(U+2 c) {\alpha^2 \over 2}+\makebox{tr}
\ln((U+g \alpha) \delta_0+c \delta_1+g\Phi)
\label{renoising}
\end{equation}
Where $\alpha=T_1$. 

To simplify this expression, we have to use the equations of motion.
Similarly as what was done in \cite{Higuchi et al},
we can express, by introducing $\Phi \rightarrow \Phi+ \epsilon
\delta_{i_n} \ldots \Phi \delta_{i_1} (z \delta_0+{c \over g} 
\delta_1+\Phi)^{-1}$ changes in variables, 
$W_0=\langle {1 \over N} \makebox{tr}
\left[z\delta_0+{c \over g} \delta_1+\Phi\right]^{-1}\rangle $
as the solution of a quartic polynomial~:
\begin{equation}
\begin{array}{l}
{\displaystyle \left( {W_0\over 2}-(z+{c\over g})
(U-g z)\right) \left( -{c \over 2 g} 
(U+2 c-g z)W_0^2 
+(2{ c^2\over g^2}+c {U\over g^2}-c {z\over g})\right. }\\
{\displaystyle  (c+g z) (U-g z)W_0+c
 (U-g z) 2 T_2 +g
 c T_{AB} \Bigg) -\left( {W_0^2 \over 4} -(z(U-g z) \right. } \\
{\displaystyle \left.+{c\over 2 g}
 (U+g z))W_0 -c+{z\over g}
(U+2 c-g z)
(c+g z)(U-g z)\right)}\\
{\displaystyle \left( {W_0^2\over 2}  -(U-g z)(z+{c\over g})W_0+2 g T_2+2 (U-g z)\right) =0}\\
\end{array}
\end{equation}

We have then to integrate :
$$\int_{{U \over g} + \alpha}^{\infty}
 \left({2 \over z}-{W_0 \over z}\left({z \over g}\right)\right)
dz$$

Without going through all the details of the procedure, let us
note that, to expand $W_0$, we have to be cautious. Indeed, as for the
one-matrix model, the integral will have to be cut into two parts, as
$U -g z$ can be large ($z \sim \infty$) or small ($z \sim {U \over g}$).
This will lead to two different changes of variables in the integral and,
once more, to logarithmic terms in $g$ \textit{and} in $- c_0$.

Let us also stress  that one has to expand the integral both in powers of
$g$ and of $c_0$. The easiest way to do so is to expand simultaneously
$I$ in powers of $g$ and $c_0$, with $g^2 \sim c_0$.
Then, we can compute approximate flows for this model.

Figure 3 shows approximate RG flows, at $x=0$, at
order $X=6$ in $g$ (i.e. three in $c_0$) and $Y=2$ in $T_2$,
where $g$ is in abcissa and $c_0$ in ordinate.

\begin{figure}[htb]
\begin{center}

\mbox{\epsfxsize=13cm \epsfbox{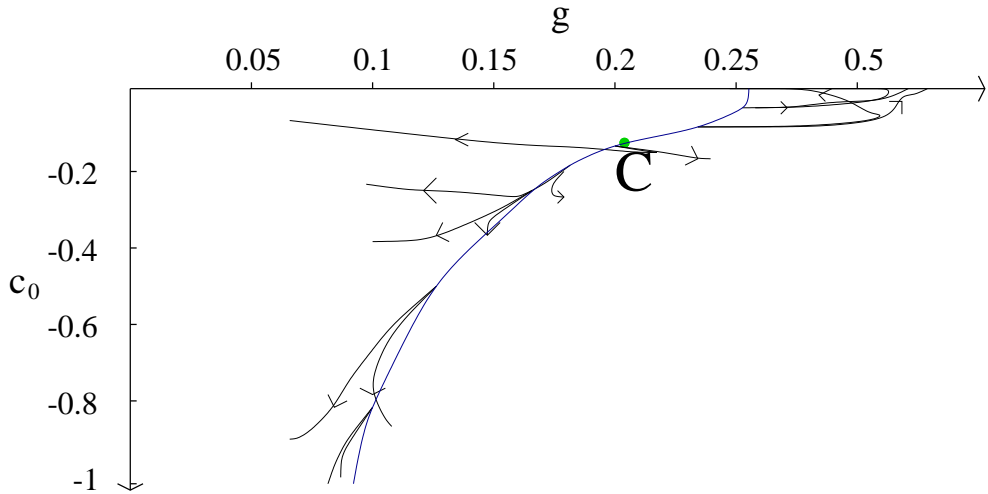}}

\caption{Approximation of the flows for the Ising model. $g$ is in abcissa, $c$ in ordinate}

\end{center}
\end{figure}

Here, we recover the pure gravity ($c_0=0$, $x=0$) critical point at 
$g_c \simeq 0.262$, i.e. with $19.5\%$ error (the exact
critical point is $432^{-1/4} \simeq 0.219$).
By extrapolating the two first results at $X=2$ and $X=3$, however,
we obtain an extrapolated $g_c \simeq 0.222$ that is to say only
$1.2\%$ error!
Moreover, we can see on figure $3$ that we recover the Ising 
bicritical point {\bf C} at $c_0 \simeq -0.14$ (the exact value is $0.15$),
and the shape of the critical line.

All these results show we can compute good approximations of the
flows, not only for one-matrix models, but also for multimatrix models.
We can theoretically compute that way the flows of an open chain of $k$ 
matrices with nearest neighbour coupling, for $k=1$, $2$(Ising), $3 \ldots$
This series of models is all the more interesting as we know that when  
$k \rightarrow \infty$, the central charge of the model $C \rightarrow 1$.
These models could thus allow us to verify the evolution of the flows
with $C$ and their shape when $C$ is equal to one.

This leads us, however, to a practical problem : though the initial model
(the open chain) has the same coupling constants for all the matrices of
the chain, the action of the renormalization group leads to almost as
many coupling constants as matrices.
This comes from the fact the roles of the matrices of the open chain are
not symmetric.
Thus, we do not know if it is practically manageable to study, for example, 
a $k=10$ matrices open chain, or if it leads to too long computations.
A solution is to study a symmetric problem : the closed chain, for
example (we do not know its exact solution), or the $k$-matrices Potts 
model, where all matrices are coupled.
The study of the latter model could be indeed very interesting :
when $k > 4$, then $C >1$ and we would thus enter the $C > 1$ domain.

\section{Conclusion}
\label{Conc}

In this paper we have developped the large $N$ renormalisation group 
method to study matrix models containing interaction terms 
corresponding to branched polymer interactions.
We have shown the analytical basis and the successes of our
method.
It can deal with models containing branched polymers, it gives us 
the shape of the flows, and also good approximations of the position of the
critical points and critical exponents of the models.
We have applied our method to the case of the pure gravity plus branched
polymer one-matrix model, and to the case of the Ising two-matrix model.
Our method is an approximation method : the exact expressions must be truncated
to a certain order to be numerically manageable (the ideal case of the infinite
chain being the exact solution).
However, the extrapolation of the first orders gives good results
without taking high computation times.
But, when studying models with a growing complexity ($k$-matrices open chains),
we may reach for big $k$ the practical limits of the method, and thus it
would be a good thing to find more technical simplifications.
We also plan to study $k$-matrices Potts models, which are very symmetric 
models (so technically simpler) but which would allow to cross the 
$C>1$ barrier for $k>4$, and are thus theoretically more complex models.

\bigskip\noindent\textbf{\large{Acknowledgements}}
\medskip

We thank J. Zinn-Justin for his interest and his careful reading of the manuscript.
This work has partial support from European contract TMR ERBFMRXCT960012.

\appendix
\section{The Vector Model}\label{VectorAppendix}
The large $N$ RG method has already been applied to the (very) simple case
of the vector model in \cite{hepth/9303090}.
This model corresponds to the limit $g=0$ of the
matrix model with action (\ref{Vt4}), since then the U($N$) matrix model
becomes a O($N^2$) vector model).
The authors of \cite{hepth/9303090} found that if one does not use the
 equations of motion (i.e. if no non-linear reparametrization
 of the fields is performed)
the RG flow equation seems to lead to strange results, with the apparent
existence of a one-parameter family of fixed points.
However they showed that when using the equations of motion one can
reduce the RG flows to much simpler flows in a finite 
dimensional space of couplings.
Moreover in that case the flows can be easily calculated exactly and it is
found that one recovers the exact critical points and critical exponents.

The purpose of this appendix is just to present the results of a simple
exercise: how do the results of the RG method change if one does not use
the equations of motion, or if one uses the equations of motion
in a approximate way?

We start from the action (\ref{Vt4}) with $g=0$ for a
 $N\times N$ Hermitean matrix
\begin{equation}
\psi(T_2)\ =\ \sum_j {h_j\over 2^{j-1}}\,{T_2}^j
\qquad , \qquad T_2\,=\ {1\over N}\,\mathrm{Tr}\left({\Phi^2\over 2}\right)
\end{equation}
Since we can rewrite $T_2=|\vec\Phi |^2$ where $\vec\Phi$ is
 the $N^2$-dimensional real vector 
whose components are the real and imaginary part of the matrix
elements $\Phi_{ij}$ of $\Phi$, the model reduces to the vector model studied
in \cite{hepth/9303090}.
The RG flow equation for the potential $\psi$ can be written exactly,
either by integrating explicitely over some components of $\vec\Phi$, or by
computing exactly the effective potential at large $N$ and
 studying its variation with $N$.
Both methods yield, to the first order in ${1 \over N}$,
\begin{equation}
N{\partial\over\partial N}\,\psi\ =\,
2\,\psi\,-\,2\,T_2\,\psi'\,+\ \ln[\psi']
\label{Vpsiflot}
\end{equation}
This simple RG equation becomes even  simpler if we consider instead of
$\psi$ its derivative $U=\psi'$ and if we invert the function $U(T_2)$ and 
consider instead the function $X=U^{-1}$ defined as $T_2=X(U)$ (this is 
at least possible for $T_2$ small).
The RG flow equation becomes linear for the function $X$

\begin{equation}
N{\partial\over\partial N}\,X\ =\ 2\,X\ -\ {1\over U}
\label{VT2flot}
\end{equation}
and is trivial to solve.
If we start at $N_0$ from the initial potential $\psi_0$, i.e. from the
initial function $X_0$ with $X_0={U_0}^{-1}={{\psi'}_0}^{-1}$
we get at $N=\Lambda N_0$
\begin{equation}
X(U;N)\ =\ {1-\Lambda^2\over 2U}\,+\,\Lambda^2\,X_0(U)\label{VXflot}
\end{equation}
and inverting again $T_2=X(U)$ we obtain the renormalized derivative
 of the potential
$U(T_2,N)={\partial\over\partial T_2}\psi(T_2,N)$.
Performing a linear rescaling $\Phi\to Z^{-1/2}\Phi$ to keep the coefficient
$h_1$ of the 
$T_2$ term fixed amounts to changing $X_0(U)\to ZX_0(ZU)$ in
\eq{VXflot}, with $Z=Z(\Lambda)$ a linear rescaling factor fine tuned such that the 
constraint
$\psi'(0)=h_1$ (i.e. $X(h_1)=0$) is kept for $N\ne 1$.

\begin{figure}[hbt]
\mbox{\epsfxsize=7.cm\epsfbox{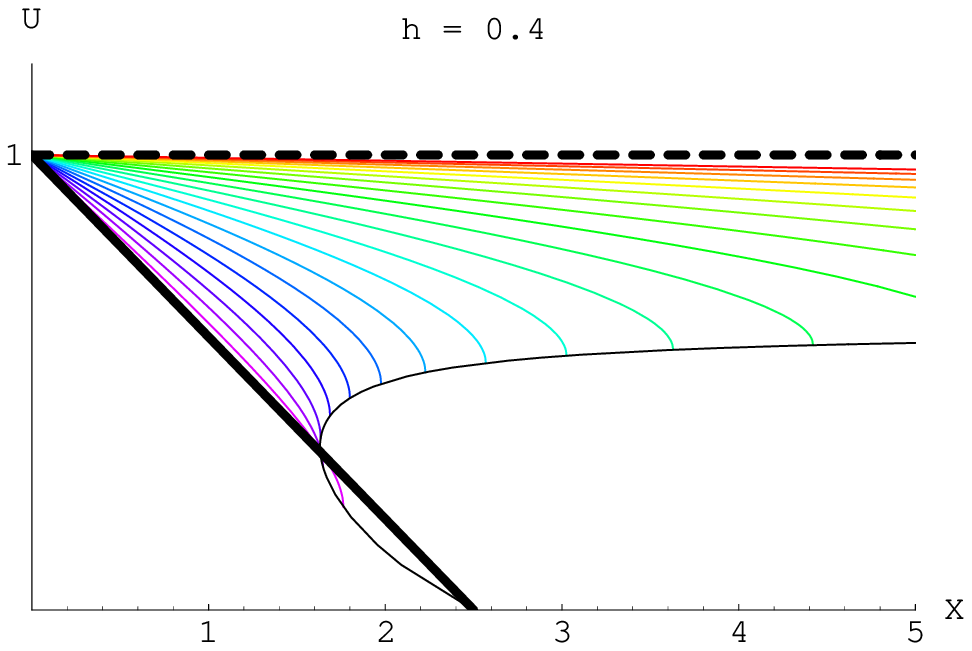}\epsfxsize=7.cm\epsfbox{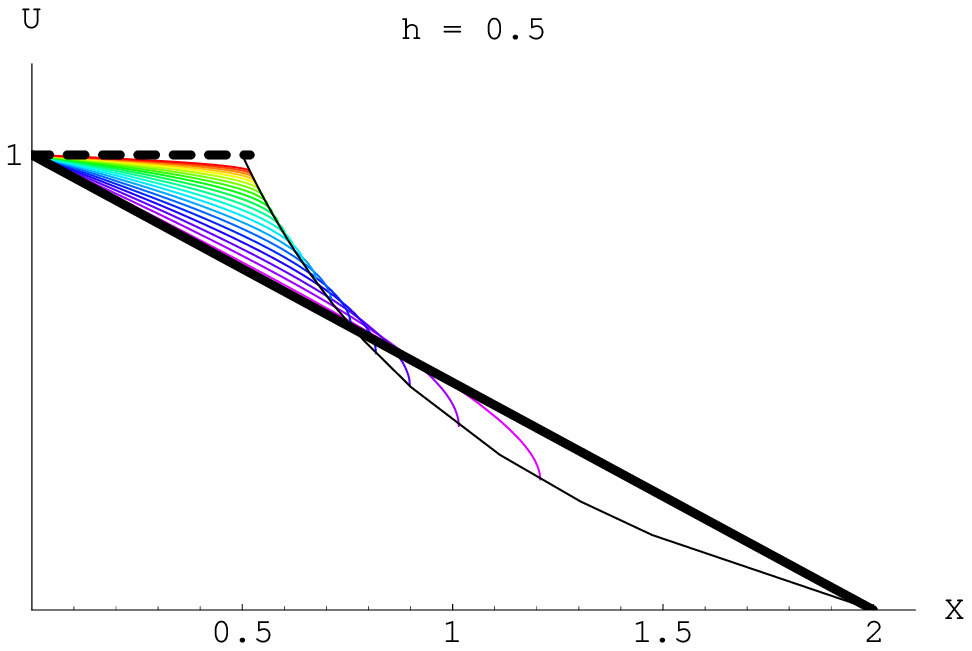}}\\
\parbox[b]{7.cm}{\epsfxsize=7.cm\epsfbox{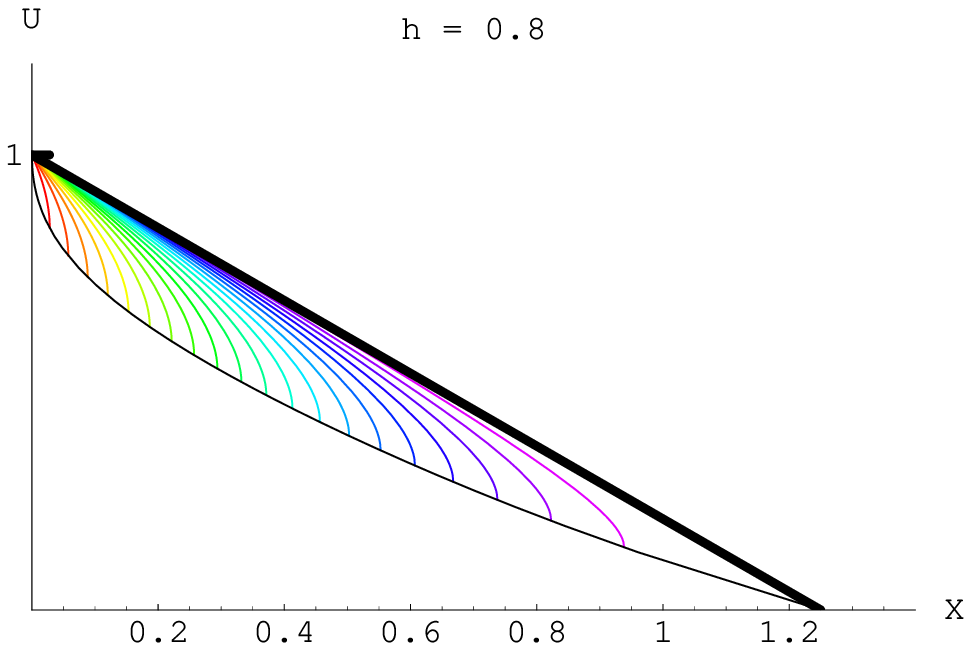}}
\raisebox{.0cm}{\parbox[b]{6.5cm}{\small{RG flows of $U=\psi'$ as
a function of $h$.
The initial function $U_0$ is the thick black line, the large $N$ limit
is the thick dashed line.
For $\Lambda>1$ $U$ develops a singularity which moves with $\Lambda$ along
the thin black curve.
The renormalized $U$ are depicted by the thin (color) curves, from
the UV with $\Lambda$ small (blue) to the IR with $\Lambda$ large (red).
for $h<h_c$ the singularity goes to $\infty$ and $U\to 1$ goes to the Gaussian
fixed point.
For $h=h_c=1/2$ $U$ goes to the non-trivial fixed point.
For $h>h_c$ the singularity hits the origin and the RG flows diverge
in a finite time.}
}}
\caption{}
\label{coloredflows}
\end{figure}

First let us study the exact flows if one starts from the quartic potential
$\psi_0=T_2-{h\over 2}{T_2}^2$, i.e. from the initial function $X_0=(1-U)/h$.
For $\Lambda>1$ small, the function $U$ remains analytic around the origin, but
it develops a square root singularity at a finite $X=X_s(\Lambda)$, which starts
for $\Lambda=1$ at the zero $X=h^{-1}$ of $U_0$, i.e. at the critical point of $\psi_0$.
Of course the flow can be studied analytically but they are better depicted
graphically (see Fig.~\ref{coloredflows})
\begin{itemize}
\item
If $0<h<h_c=1/2$, the singularity $X_s(N)$ goes to infinity as $N\to\infty$ and
the function $U(X,N)$ tends towards the constant function $U(X)=1$,
which corresponds to the Gaussian potential $\psi=T_2$
(Gaussian fixed point) (Fig.~\ref{coloredflows}.a).
\item
At the critical value $h_c=1/2$ the singularity $X_s$ tends toward
a \textit{finite value} $X^*=1/2$, so that the function $U$ tends toward a 
non-analytic fixed point  (Fig.~\ref{coloredflows}.b)
\begin{equation}
U^*(X)\ =\ \left\{
\begin{array}{l}
1\ \mathrm{if}\ X\le X^*\\
\mathrm{undefined\ if\ }X>X^*
\end{array}
\right.
\end{equation}
\item
For $h>h_c=1/2$, the singularity $X_s(N)$ reaches the origin $X=0$ in a
\textit{finite} RG time $\Lambda_s=2h/(2h-1)$, the potential $\psi$ becomes
 singular at
the origin and the RG flow thus diverges in a finite time and reaches no
fixed point (Fig.~\ref{coloredflows}.c). 
\end{itemize}
This analysis can be extended to general initial potentials.
We thus recover (without using the equations of motion)
a sensible picture of the RG flows:
the attraction domain of the 
Gaussian fixed point, ($h<h_c$) is separated from the domain where RG flows
diverge by a critical (unstable) manifold where one is driven
towards the non-trivial fixed point $U^*$ (which corresponds to
branched polymers).
The only subtle point is that the non trivial fixed point is non-analytic
and cannot be distinguished by a local analysis around the origin ($T_2=0$)
from the Gaussian fixed point.
This explains the apparent paradoxes of \cite{hepth/9303090}.

Let us mention that if instead of using the potential $U(X)$ one uses the
inverse function $Y(U)=X(U)-1/2U$, the RG flot is very simple: 
$Y(U)\to \Lambda^2 Y(U)$.
The structure of the flow is described by the expansion of the function
$Y(U)$ around the largest zero $U^*$ such that $Y(U^*)=0$, and the (only) relevant scaling field $t_0$ corresponds to the first derivative $Y'(U^*)$ of
$Y$ (since the critical manifold corresponds to $t_0=0$).
This $t_0$ scales with $\Lambda$ as $\Lambda^2$, therefore has scaling dimension $2$.
However the mapping $U(X) \to t_0$ is highly non-linear, and becomes
singular along the critical manifold $t_0=0$.
Therefore this does not contradict the fact that the real scaling dimension
of the relevant operator is $4/3$.

%Let us mention that if one parametrises the potential $U(X)$ by the
%expansion of X(U) around the
%singularity at $X_s=1/2$ (this requires using changes of scaling variable which
%become singular in the limit $N\to\infty$) the RG flow equation becomes
%regular and can be linearized in the vicinity of the fixed point.
%One recovers in that way the correct scaling fields and scaling dimensions,
%in particular the dimension for the relevant operator $\lambda_0=4/3$.

One can try to truncate the RG flow equation \eq{Vpsiflot}
 in the most naive way:
we keep only the couplings $h_j$ with dimension $j\le K$ in $\psi$, then
expand the $\ln[\psi']$ in powers of $T_2$ and {\it truncate this expansion
at order K in ${T_2}$}.
We thus obtain for fixed $K$ approximate RG flow equations for the
$K$ couplings $h_j$,
which are of the standard form ${\scriptstyle N}
{\partial\over\partial N}h_j=\beta_j[h]$, with the
 $\beta$ functions polynomials of order $K$ in the
$h_j$'s.
At a given truncation order $K$ the explicit form of these functions is not 
especially illuminating and will not be given here. 
Using computing software the approximate
 fixed points can be found exactly and
the structure of the RG flows studied.
We find indeed a non-trivial fixed point $\psi^*$ with one unstable
direction, which should correspond to the branched polymer fixed point.

\begin{figure}[bht]
\begin{center}
\mbox{\epsfbox{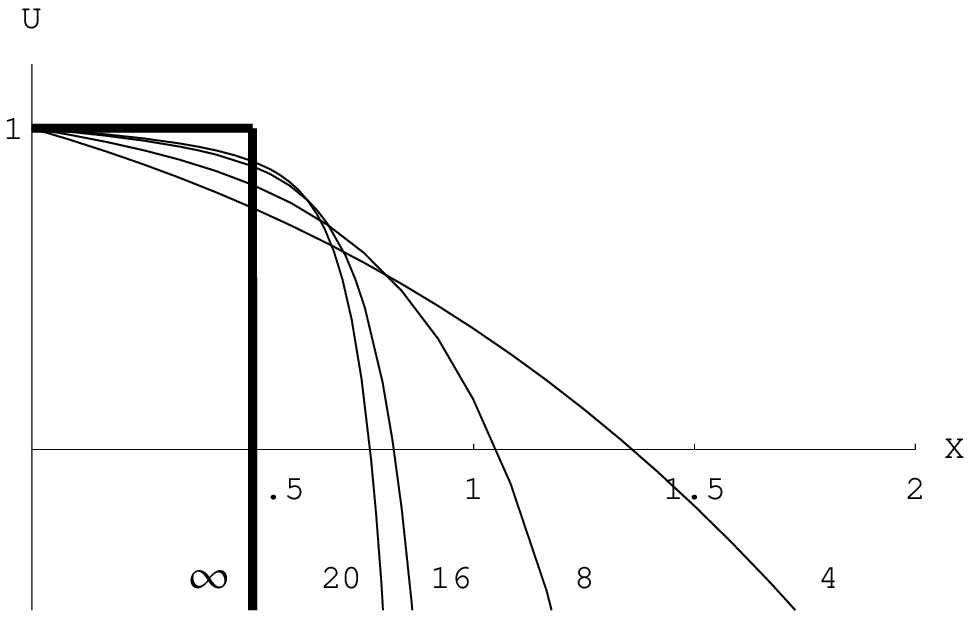}}
\end{center}
\caption{}
\end{figure}

The derivative $U^*={\psi^*}'$ is depicted on Fig.~\ref{coloredflows}
as a function of the order of truncation $K$ ($U$  is a polynomial of degree
$K-1$.
One sees that as $K$ increase the approximated fixed points converge towards
the exact (but singular) fixed point.
However a more precise analysis (that we do not reproduce here) shows that
the convergence is very slow, typically as $1/K$.
The finite $K$ estimates for the critical coupling $h_c$ 
(if all higher order couplings $h_j$, $j>2$ are set to zero) 
converge towards the exact value $1/2$ at the same rate.

This very slow convergence is insufficient to obtain good estimates for
the critical exponents. 
It turns out that within this approximation scheme, the scaling dimension
$\lambda_i^{\mathrm{appr}}$
of the scaling fields at the approximate fixed point are independent of
the trucation order $K$!
Indeed they are found to be integers $(1,-1,-2,-3,\ldots)$.
The $+1$ dimension should correspond to the dimension
 $\lambda_0$ of the relevant perturbation, which is known to be $4/3$.
Thus, although the estimates for the critical points converge towards the
correct result as $K\to\infty$, the estimates for the scaling exponents
do not!
A procedure to accelerate the convergence is required.
This is precisely what the equation of motions are doing.

\end{document}